# Directional Wetting of Submerged Gas-entrapping Microtextures

Sankara Arunachalam[*] & Himanshu Mishra[*]

[1]Biological and Environmental Sciences and Engineering Division, King Abdullah University of Science and Technology (KAUST), Thuwal 23955-6900, Kingdom of Saudi Arabia

[2]Water Desalination and Reuse Center, KAUST

[*]Correspondence: sankara.arunachalam@kaust.edu.sa & himanshu.mishra@kaust.edu.sa

**Abstract**

Numerous natural and industrial processes entail the spontaneous entrapment of gas/air as rough/patterned surfaces are submerged under water. As the wetting transitions ensue, the gas diffuses into the water leading to the fully-water-filled state. However, the standard models for wetting do not account for the microtexture's topography. In other words, it is not clear whether the lifetime of n cavities arranged in a I-D line or a II-D (circular or square) lattice would be the same or not as a single 0-D cavity. In response, we tracked the time-dependent fates of gas pockets trapped in I-D and II-D lattices and compared them with wetting transitions in commensurate 0-D cavities. Interestingly, the wetting transitions in the I-D and the II-D arrays had a directionality such that the gas from the outermost cavities was lost the first, while the innermost got filled by water the last. In essence, microtexture's spatial organization afforded shielding to the loss of the gas from the innermost cavities, which we probed as a function of the microtexture's pitch, surface density, dimensionality, and hydrostatic pressure. These findings advance our knowledge of wetting transitions in microtextures, and inspiring surface textures to protect electronic devices against liquid ingression.

**Introduction**

Air entrapment at the liquid–solid interface can be a boon or a bane depending on the context [1], e.g., underwater plastron breathing [2], waterproof electronic devices [3], ultrasound contrast agents [4], and cavitation damage in plants [5], machinery, and human embolism [6]. Whilst the time-dependent dissolution of isolated bubbles in bulk water and surface-bound bubbles has been studied [7, 8], the collective dissolution of entrapped air pockets/bubbles has been lacking. Recently *Michelin* et al. employed numerical simulations to probe microbubbles dispersed in undersaturated liquids and found a diffusive shielding effect in their dissolution, i.e., the bubbles in the interior dissolved slower than those at the boundary [9]. Previously, Lohse et al., probed coarsening surface-bound nanobubbles [10, 11] and Zhang et al., also presented the collective bubble growth on microtextured surfaces spurred via solvent exchange has also been investigated [12]. However, the significance of neighboring/interacting gas pockets on the wetting transitions on surfaces comprising of microcavities has been largely unexplored. Remarkably, such air-filled metastable Cassie states may transition to the fully-filled Wenzel state during the course of seconds, hours, days, weeks, months, and years depending on the liquid vapor pressure, condensation, apparent contact angles, and gas diffusion [13-15]. For instance, it is not entirely clear if surface-bound gas pockets arranged in one-dimensional (I-D) or one-dimensional (II-D) arrays would all dissolve at the same rate (or faster or slower) in comparison with an isolated cavity (0-D) under identical conditions.

To elucidate this gap, we use gas-entrapping microtextured surfaces comprising simple cylindrical cavities and present the time-dependent collective wetting pattern in 0-D (singly cavity), I-D (line distribution), and II-D (circular and square arrays). Specifically, we address the following interrelated elemental questions pertaining to the directionality of wetting transitions: (i) How does the array's dimensionality, viz. 0-D (individual cavity), I-D (a line of cavities), or II-D arrays influence the lifetime of underwater air entrapment? (ii) For a fixed number of cavities, how do I-D and II-D lattices compare in terms of longevity enhancement? (iii) How do different II-D arrays, e.g., circular or square, impact directionality?

**Results and Discussion**

Using photolithography and dry etching protocols [16] (Methods), we realized microtextured surfaces comprised of cylindrical cavities in 0-, I-, and II-D arrays onto commercial $SiO_2/Si$ wafers(Figs. 1A-E). The diameter, depth, and pitch (center-to-center distance) of the cavities were set to $D = 200$ µm, $h \approx 58$ µm, and $L = 220$ µm (Fig. S1). To facilitate air entrapment under water, all the microtextured surfaces were grafted with perfluorodecyltrichlorosilane (FDTS). Consequently, the advancing and receding contact angles of water droplets on FDTS-coated flat $SiO_2$ were $\theta_A = 120° \pm 1°$ and $\theta_R = 100° \pm 2°$. Next, using a custom-built pressure cell, we (horizontally) submerged the samples under stagnant water columns of height $H = 6$ mm. To observe the wetting transitions from the top, a white light microscope with a camera was mounted to the cell (Fig. 1F, S2, Methods).

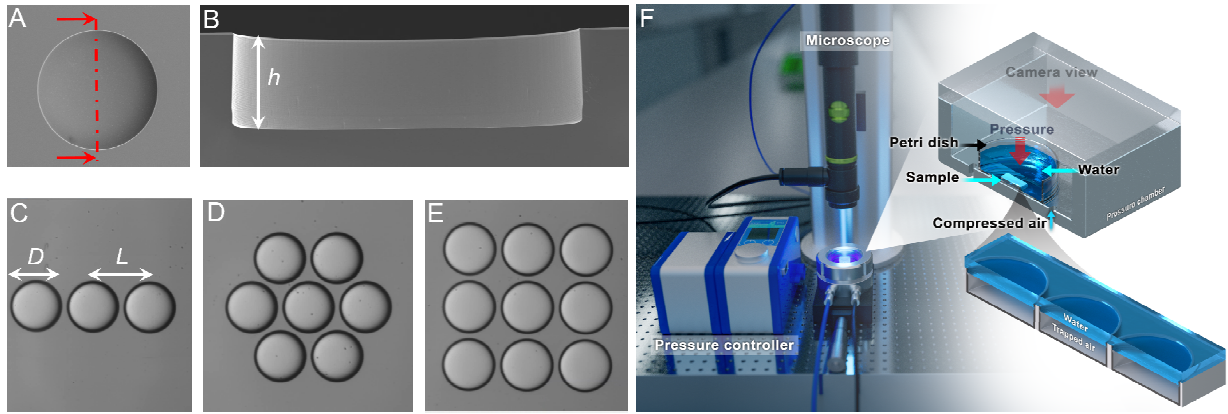

**Fig. 1.** Representative scanning electron micrographs of microfabricated surfaces. (**A-B**) top and cross-sectional view of a single cavity, (**C-E**) array of cavities arranged in a I-D (line) or II-D (circular or square) lattice (**F**) Schematic of the experimental setup to study pressure-induced wetting transitions. For all the cases, cavity diameter = 200 µm, pitch = 220 µm.

To expedite wetting transitions, we pressurized the headspace air, at the rate of 1 kPa/s to reach 1.20 atm (absolute pressure). At this pressure, the water meniscus first drooped inside the 0-D (single) cavity and touched its floor in $t_0 = 5.9 \pm 0.1$ min (fail time); the fully-filled (or the Wenzel) state was reached in a total of 11 mins (Fig. 2A). According to Henry's law, pressurizing the water above the atmospheric pressure makes it undersaturated with respect to the trapped bubble. As gas dissolves in the water, the water meniscus enters the cavity, and the Cassie-to-Wenzel wetting transition ensues.

**Wetting Directionality in I-D Lattice:** Interestingly, as an additional cavity is added next to the 0-D case, the failing time of both the cavities is extended by ≈ 20% (Fig. 2B). For $n \geq 3$, it was apparent that the cavities at the either edge of the I-D lattice failed first and the innermost cavity was the last to fail (Fig. 2C-F, Movie S1-S2). For instance, as the number of cavities was increased in the range of $n = 3$–9, the failure of the innermost cavity was delayed by 60–130% in comparison to the 0-D case under the same conditions (Fig. 2C-F).

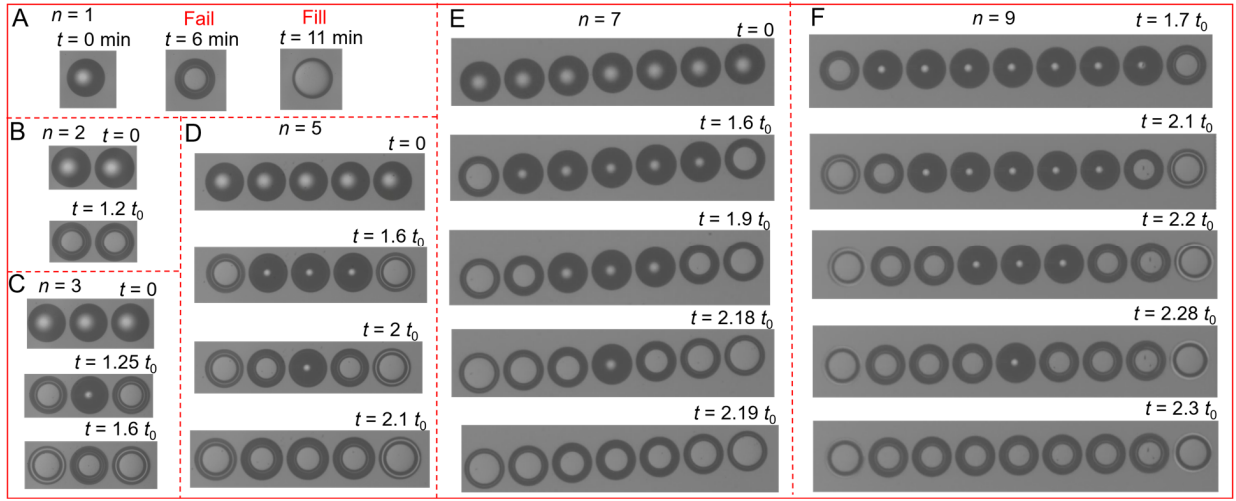

**Fig. 2.** Directional wetting transitions in 0-, I-D lattices. Optical micrographs of silica surface with arrays of simple cavities immersed under a 6 mm water column and headspace air pressure = 1.2 atm. The fail time for $n \geq 2$ cases were higher than the commensurate single cavity; the failing front also propagates from the extremal least-shielded cavities toward the center cavity. For all the cases, cavity diameter = 200 µm, pitch = 220 µm.

**Wetting Directionality in II-D Lattices.** To investigate wetting directionality in II-D lattices, the cavities were organized in circular or square lattices and wetting transitions were observed under 1.2 atm. In the circular lattice, the outermost cavity layer failed the first, which was followed by the consecutive inner layers (Fig. 3AB, Movie S3). Since the inner cavities were shielded up until all the outer cavities failed, the time for the centermost cavity to fail increased significantly (Fig. 3AB). In other words, the shielding effect was the highest at the center and the lowest on the outskirts of the array. As the number of the layers increased from $N = 2$, 3 and 4 (i.e., total number of cavities, $n = 7$, 19 and 37), the failure time of the innermost cavity increased by 310%, 510% and 790%, compared to the 0-D case (Movie S3). In the square lattice, the corner cavities failed the first, followed by the cavities located in the outermost layer and this trend repeated until the centermost cavity was reached (Fig. 3CD Movie S4). As the number of layers increased from $N =$ 2 to 3 (and $n = 9$ and 25), the failure time of the innermost cavity was increased by 410% and 650%. Interestingly, if the organization of 7 (or 9) cavities is changed from I-D to II-D, the failure time of the innermost increases from 120% (or 130%) to 310% (or 410%) (Fig. 2EF and Fig. 3AC).

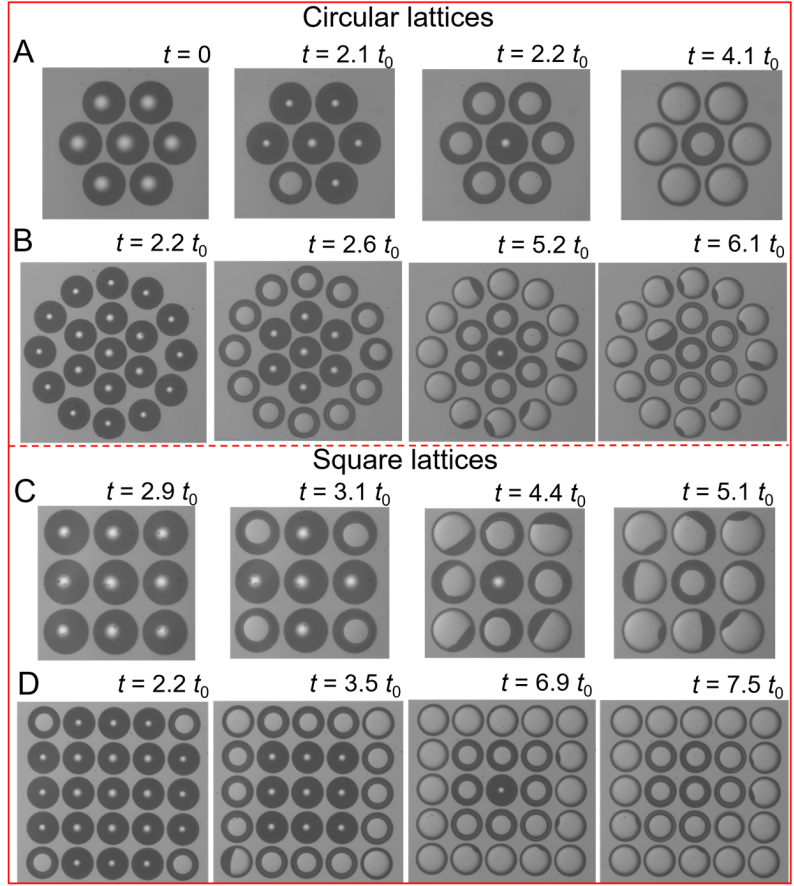

**Fig. 3.** Directional wetting transitions II-D lattices. Optical micrographs of silica surfaces with circular and square arrays of simple cavities immersed under a 6 mm water column and headspace air pressure = 1.2 atm. In circular lattices, the outermost cavity layer failed the first, which was followed by the consecutive inner layers. In square lattices, cavities located at the corners fail the first, followed by the cavities at the outermost boundary, and this trend continues inward. For all the cases, cavity diameter = 200 µm, pitch = 220 µm.

Here, we explain qualitatively the factors and mechanisms underlying the shielding effect and directional wetting transitions in arrays of cavities submerged under water. The dissolution of the trapped air under hydrostatic pressure causes the primary wetting transition. According to Henry's law ($P_a = K_H c_o$, where $K_H$ is Henry's law constant) as we increase the air pressure ($P_a$), above the liquid column, from 0 to 20 kPa, the equilibrium air-water interface concentration increases, $c_\infty = 0.023$ kg/m³ to $c_o = 0.028$ Kg/m³ [17, 18]. This concentration gradient ($c_o$ – $c_\infty$) drives the outward gas diffusion at the speed of diffusivity, $D_a = 2\times10^{-5}$ cm$^2$/s through the gas-liquid interface area. The typical bubble dissolution time scale is $t_{dis} \approx \frac{\rho_\infty}{(C_0 - C_\infty)} t_{dif}$, where $t_{dif} \approx \frac{l^2}{D_a}$ is the typical diffusion time scale, $\rho_\infty$ is the density of the air, $l$ is the length of the diffusion [10]. Our experimental duration for a single cavity to reach a fail state under 20 kPa is 6 min, resulting in a diffusion length of 848 µm. As a result, the gas dissolving from the cavities arranged in a I- or a II-D lattice diffuses normally and laterally, enhancing the background aqueous gas concentration over the neighboring cavities due to the diffusional interaction. Consequently, the loss of the entrapped gas from the interior cavities slows down, while the boundary cavities with lesser neighbours experience a significantly higher diffusional loss in comparison. The outer/corner cavities are the first to fail, albeit the failure times are still longer than those of the 0-D cases (Fig. 4).

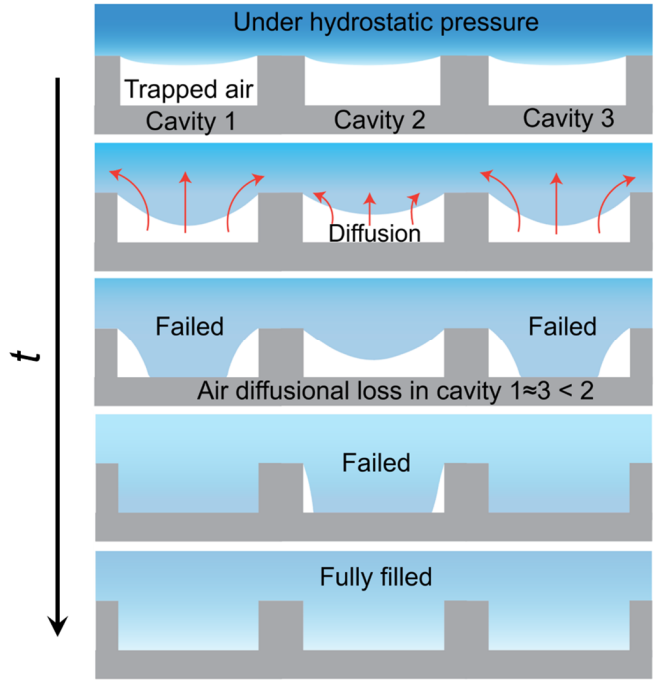

**Fig. 4.** Illustration of shielding effect induced wetting directionality. As the pressure increases, each cavity acts as a source of dissolved gas for its neighbor, raising the background concentration seen by each gas cavity and slowing down its dissolution. Since the boundary cavity has no neighboring cavities, it experiences more diffusional loss and fails first. This difference in diffusional air loss from individual cavity induces directionality.

The spatial (I-, or II-D) organization of the cavities affords a "shielding effect" to the failure of the inner cavities. The delayed wetting transitions due to the shielding effect can be expressed as $(t_a^{max}/t_0) = 1 + \widetilde{t_{SE}}$, where $t_a^{max}$ is the maximum fail time of cavity in an array, $t_0$ is the fail time of single cavity, and $\widetilde{t_{SE}}$ is the delay due to the shielding effect. Remarkably, the $\widetilde{t_{SE}}$ values for the I-D and II-D lattices scale as $(log\ n)/L^*$ and $\sqrt{n}/L^*$, respectively, where $n$ is the number of the cavities and $L^*$ is the non-dimensional pitch ($L^* = L/D$) (Fig. 5A) [9]. These trends are in agreement with the predictions of Michelin et al., for the collective dissolution of gas bubbles in undersaturated liquid [9].

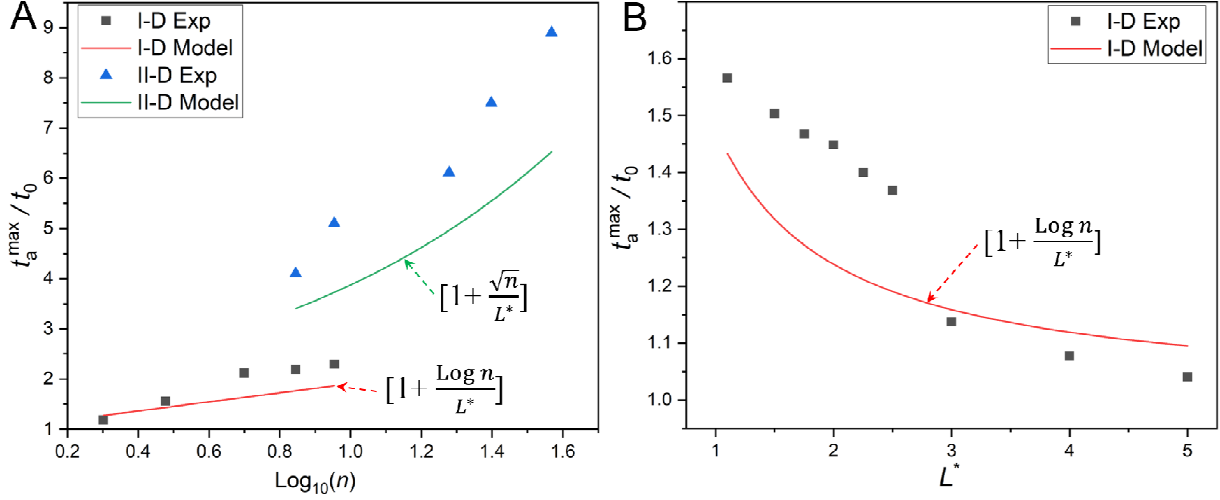

**Fig. 5.** Experiment versus model prediction. (A) Extended fail time due to shielding effect in I- and II-D lattices. (B) Effect of pitch on the extended fail time of three cavities distributed in I-D.

To pinpoint the geometrical arrangements at which the shielding effect is no longer effective, at enhancing $\widetilde{t_{SE}}$ of the innermost cavities, we increased the pitch to $L = 5D$ for the I-D case and to $6D$ for the II-D case. While the shielding-induced directionality of cavity filling was still observed, $\widetilde{t_{SE}}$ approached to zero (Fig. 5B, Fig. S3 and Movie S5). A similar situation arises when increasing the applied pressure while keeping the same pitch. For instance, failure time of the centermost cavity in square lattices with $N = 3$ layers submerged under 50 kPa was reduced to 400% (Fig. S4, Movie S6). These findings reveal that as the microtexture pitch $L$ is lower than the diffusion length pertaining to the lifetime of the 0-D case, $\sqrt{D_a t_0}$, the shielding effect will dominate and wetting transitions in the innermost cavities will be delayed.

**Conclusion**

To summarize, we reported on a hitherto unappreciated shielding effect that could be exploited to tune wetting transitions in specific (innermost) regions of gas-entrapping microtextured surfaces submerged in undersaturated liquids. These insights should be useful in understanding the diffusional interaction of collective bubbles/droplets [1, 11, 19] and gas exchange from the plastrons of underwater insects [2]; as well, their potential should be explored to engender robustness in perfluorocarbon-free gas-entrapping microtexutres against accidental liquid ingression in smart devices [3, 15]. Many of the fundamental insights gained here can be applied to other liquids (e.g., oils), contact angles, and cavities or pores of different dimensions or geometries.

**Materials and Methods**

**Microfabrication**: Surfaces adorned with arrays of microscale cylindrical cavities were created using standard photolithography and dry etching procedures on a silicon wafer (4-inch diameter, <100> orientation) with a 2 µm thick silica layer on top. The cavity pattern was prepared using Tanner EDA L-Edit software and then transferred to the AZ5214E spin-coated wafer using a Heidelberg Instruments µPG501 direct-writing device. The UV-exposed photoresist was removed in a bath of AZ-726 developer. The exposed $SiO_2$ top layer was etched away in an Inductively Coupled Plasma (ICP) Reactive-Ion Etching (RIE) equipment before being moved to an Oxford Instruments Deep ICP-RIE to etch the Si beneath the $SiO_2$ layer. After microfabrication, we cleaned the wafers in piranha solution comprising of sulfuric acid ($H_2SO_4$, 96%): hydrogen peroxide ($H_2O_2$, 30%) = 3:1 (v/v) maintained at $T = 388$ K for 10 min and spin dried under Nitrogen environment.

**Scanning electron microscopy:** Selected samples were cleaved using a diamond tip scribe and coated with a 3 nm Iridium layer before being observed by Quattro SEM at 5 kV accelerating voltage and 28 pA current.

**Silanization procedure:** Using the ASMT Molecular Vapor Deposition (MVD) 100E equipment, we chemically grafted gas-phase Perfluorodecyltrichlorosilane molecules (FDTS) onto our silica surfaces.

**Advancing/receding contact angles in air:** The static and advancing/receding contact angle measurement using de-ionized water were performed in a Kruss Drop Shape Analyzer - DSA100 at 0.2 µL s$^{-1}$. All the data were analyzed using the *Advance* software. Reported data points are an average of 3 measurements.

**Pressure experiments:** We used a custom-built pressure cell along with a fluigent pressure regulator and compressed air to increase the headspace pressure on top of the water which eventually increases the hydrostatic pressure. The pressure ramp rate was set as 1 kPa/s. Microcavities were observed from the top using an optical microscope connected to USB camera. We fixed the sample into the petri dish (35 mm petri dish from Celltreat) using double-sided tape (Scotch) and kept it inside the cell, before the experiment, we added the known amount of water (≈ 6 ml at ≈ 0.3 ml/s rate) using a syringe and needle to submerge the sample and applied external pressure. All the experiments were carried out at room temperature 20°C.

**Supplementary Information** is available in the on-line version of this paper.

**ACKNOWLEDGEMENTS.** SM thanks Dr. Subkhi Sadullah and Mr. Zain Ahmad for the fruitful technical discussions.

**Author Contributions:** S.A. conceived the idea, performed the research under H.M.'s supervision, and jointly wrote the paper.

**Conflict of Interest:** The authors have no conflicts to disclose

**DATA AVAILABILITY.** The data that support the findings of this study are available from the corresponding author upon reasonable request.

# Supporting Information for

Directional Wetting of Submerged Gas-entrapping Microtextures

Sankara Arunachalam & Himanshu Mishra

[1]Biological and Environmental Sciences and Engineering, Division, King Abdullah University of Science and Technology (KAUST), Thuwal 23955-6900, Kingdom of Saudi Arabia

[2]Water Desalination and Reuse Center, KAUST

[*]Correspondence: sankara.arunachalam@kaust.edu.sa & himanshu.mishra@kaust.edu.sa

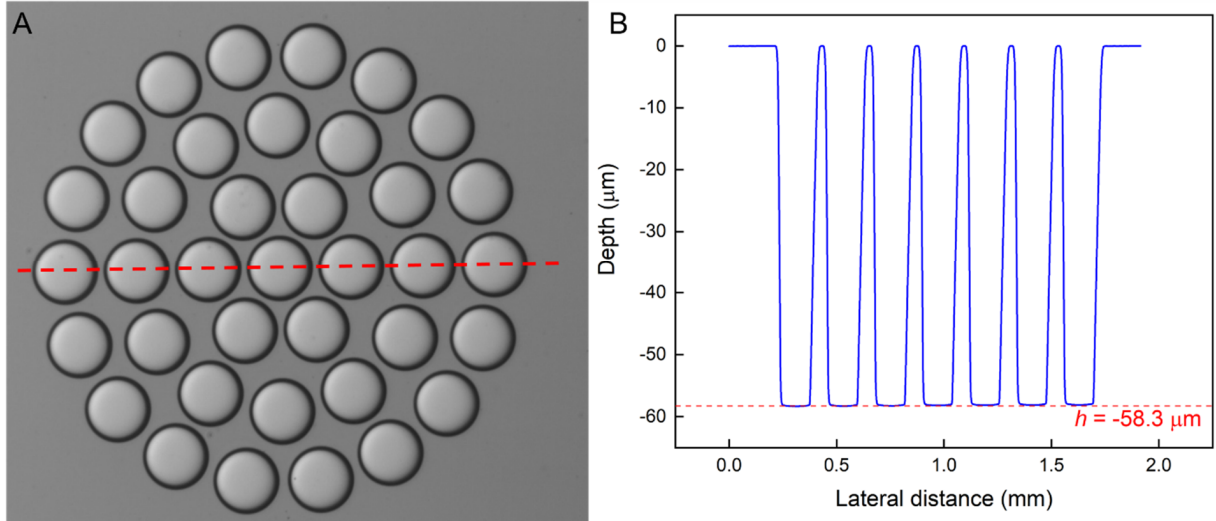

**Fig. S1.** Depth uniformity measurement for SCs. Cavity depth was measured along the dotted line using a profilometer (DektakXT - Bruker), and it looks identical.

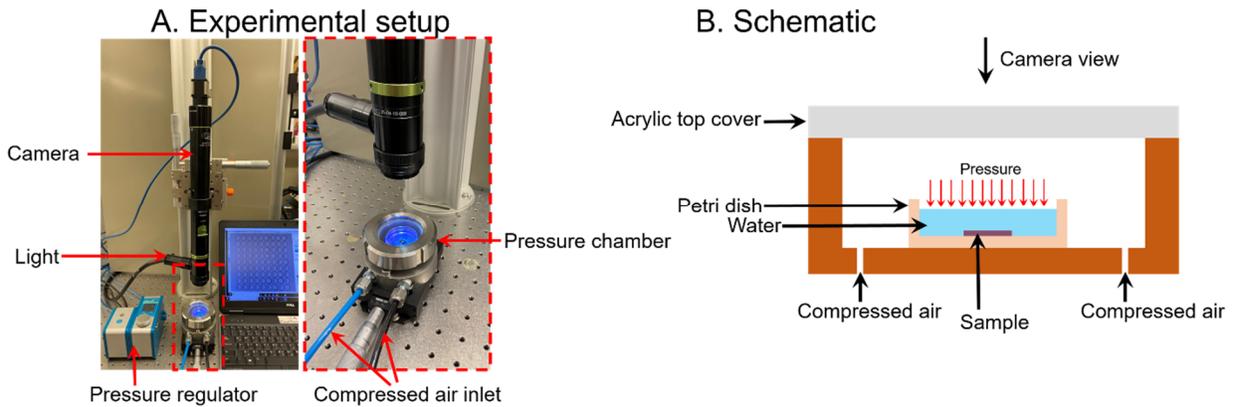

**Fig. S2**. (**A**) Experimental setup to determine failing directionality of the cavities under elevated pressure. Optical images were recorded using a vertically mounted USB camera. (**B**) Samples were immersed in water inside the pressure cell. Subsequently, we used compressed air to apply external pressure controlled by a regulator.

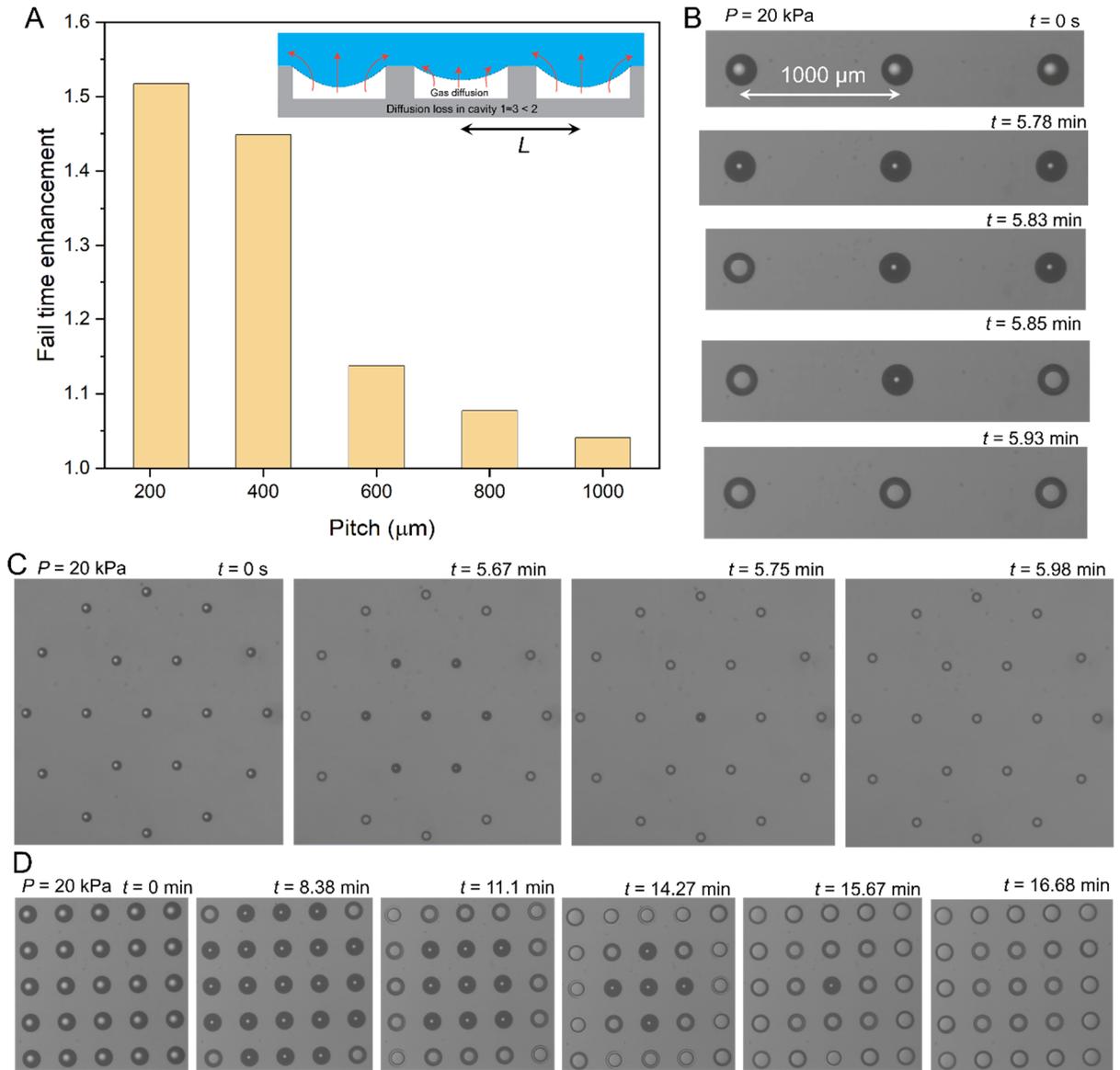

**Fig. S3.** Effect of inter-cavity distance on directionality in ID and IID distribution. (A) Three cavities with various inter cavity distance (B) Circular lattice, $N = 3$ and pitch 1200 µm, directionality but shielding effect reduced drastically. C. Square lattice, $N = 3$, pitch $L = 400$ µm. Symmetric inward filling of cavities corner and then boundary cavities followed by a central cavity. For all the cases cavity diameter is 200 µm, immersed under a 6 mm water column, and applied air pressure = 20 kPa.

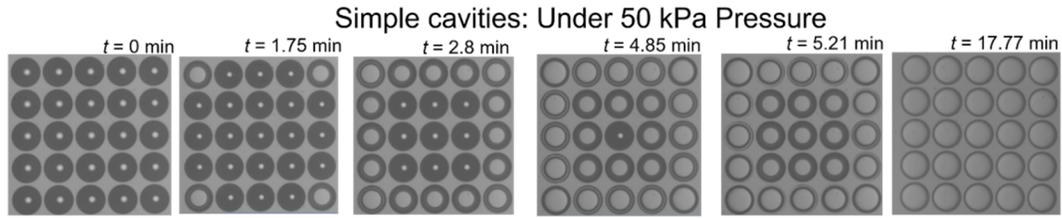

**Fig. S4**. Directional wetting of II-D square lattices under 50 kPa pressure. The cavities located at the corners fail the first followed by the cavities at the outermost boundary and this trend continues inward. Scale bar: the cavity diameter is 200 µm.

**Movie captions:**

**Movie S1.** Wetting transitions on microtextured silica surfaces with simple cylindrical cavities. Surfaces with line of simple cavities ($n$ = 1,2 and 3), immersed under a 6 mm water column, headspace air pressure = 1.2 atm. For all the cases, cavity diameter = 200 µm, pitch = 220 µm. Movie playback speed: 20x faster.

**Movie S2.** Directional wetting transitions in I-D, line distributions. Microtextured silica surfaces with a line of simple cylindrical cavities ($n$ = 5 and 9), immersed under a 6 mm water column, headspace air pressure = 1.2 atm. For all the cases, cavity diameter = 200 µm, pitch = 220 µm. Movie playback speed: 20x faster.

**Movie S3.** Directional wetting transitions in II-D, circular lattice. Microtextured silica surfaces with simple cylindrical cavities arranged in a circular lattice, $N$ = 2, 3 and 4. Samples were immersed under a 6 mm water column, headspace air pressure = 1.2 atm. For all the cases, cavity diameter = 200 µm, pitch = 220 µm. Movie playback speed: 120x faster.

**Movie S4.** Directional wetting transitions in II-D, square lattice. Microtextured silica surfaces with simple cavities arranged in a square lattice, $N$ = 2 and 3. Samples were immersed under a 6 mm water column, headspace air pressure = 1.2 atm. For all the cases, cavity diameter = 200 µm, pitch = 220 µm. Movie playback speed: 60x faster.

**Movie S5.** Effect of inter-cavity distance on directional wetting transitions in I- and II-D. Microtextured silica surfaces with simple cylindrical cavities (i) line of cavities, $n$ = 3 and pitch, $L$ = 1000 µm (ii) circular lattice, $N$ = 3 and $L$ = 1200 µm (iii) square lattice, $N$ = 3, $L$ = 400 µm. Samples were immersed under a 6 mm water column and headspace air pressure = 1.2 atm. For all the cases cavity diameter is 200 µm. Movie playback speed: 30x faster.

**Movie S6.** Effect of hydrostatic pressure on directional wetting transitions in II-D, square lattice. FDTS-coated microstructured silica surfaces with simple cylindrical cavities arranged in a square lattice, $N$ = 3. The sample is immersed under a 6 mm water column and applied air pressure = 50 kPa. Cavity diameter = 200 µm, pitch = 220 µm. Movie playback speed: 20x faster.